# Characterization of a supersonic gas jet via laser-induced photoelectron ionization

Alexander Golombek, Lisa Danzig and Andreas Wucher*

*Fakultät für Physik, Universität Duisburg-Essen, 47048 Duisburg, Germany*

*email: andreas.wucher@uni-due.de

## Abstract

We describe the characterization of a pulsed supersonic rare gas beam which is intended to serve as an ultracold neutral atom target for the production of an ultrashort ion pulse via femtosecond photoionization. The velocity distribution of atoms entrained in the beam is measured and characterized by temperatures $T_\parallel$ and $T_\perp$ in directions along and perpendicular to the beam propagation, respectively. It is shown that $T_\perp$ values in the mK regime are achieved at distances of the order of 1000 mm from the nozzle. Moreover, the center beam density at this position is measured to be of the order of $10^{11}$ atoms/cm³. Both findings are essential for the intended application and confirm the targeted beam specifications. Comparison with theoretical estimates reveals the well-known skimmer interference effect, which is found to reduce the beam density by more than one order of magnitude.

## 1. Introduction

Supersonic gas expansions [1] of atoms or molecules play a huge role in different areas in physics as well as physical chemistry, e.g. matter-wave microscopy [2], ultrahigh-resolution spectroscopy [3] and laser-induced fluorescence spectroscopy [4]. The increasing emergence of new experimental applications is mostly due to the risingly sophisticated creation and control as well as characterization of molecular gas jets. Moreover, computer simulations [5, 6] allow to study more precisely the influence of nozzle and skimmer shapes on the beam properties, which facilitates the production of such components for an experimental implementation enormously.

A supersonic gas expansion is created by leading gas atoms or molecules from a reservoir with high pressure through a small nozzle exit against an evacuated chamber. This is frequently done in a pulsed manner, thus generating single gas pulses with a defined spatial extension along its expansion axis. A skimmer with a given diameter inserted into the beam line allows to collimate the beam to only the central part of the beam and mostly defines the spatial shape of the beam profile at a given plane perpendicular to the beam axis. While propagating alongside its center line, the temperature of a supersonic gas jet is rapidly decreasing with increasing distance from the nozzle. Typically, after a few nozzle diameters the trajectories of the atoms are considered to be "frozen out", i.e. the particles rarely interact with each other, resulting in a gas pulse with minimal relative movement of the particles. This movement is commonly described with a Boltzmann distribution for the velocities parallel ($v_\parallel$) and perpendicular ($v_\perp$) to the propagation direction and associated with temperatures $T_\parallel$ and $T_\perp$

, respectively [7]. While contributions determining $T_\perp$ have proven to be difficult, measurements of $T_\parallel$ have been done extensively in the past [8]. In doing so, the parallel temperature $T_\parallel$ typically lies in the order of 0.1 to a few Kelvin while the perpendicular temperature $T_\perp$ should be substantially lower, as predicted by theoretical considerations [9].

Another important property of the beam is its center line density, which decreases with increasing distance from the nozzle and can be estimated theoretically from the initial nozzle parameters [9]. Nevertheless, the theoretically expected density is rarely achieved in experiment due to the presence of skimmers, which cause particles to scatter from the aperture edges, thereby increasing the beam divergence as well as the internal temperature of the beam. For experiments like molecular cooling [10] or trapping [11], however, the knowledge of the beam density is crucial, therefore techniques were developed to measure the number density of particles within the expanding beam. Most of the detection techniques base upon indirect measurements such as laser induced fluorescence [12] or laser interferometry [13]. Recently, various research groups showed a direct approach to measure the properties of a supersonic gas expansion via photoionization using fs laser pulses [14] in combination with fs ion imaging [15]. Those direct measurements constitute a promising way to investigate the characteristics of a supersonic gas expansion, as the direct approach has substantial advantages. The laser beam can be tightly focused, thereby producing photoions which can be easily controlled with electrical fields and detected with standard detectors such as a multichannel plate. In combination with a time-of-flight analysis, not only the total particle density but also the particle composition within the supersonic beam can be determined. By scanning the laser through the beam, this information can be obtained in a spatially resolved manner. Moreover, a quantitative calibration of the beam density can be obtained by comparing the resulting photoion signal to that measured via backfilling the chamber to a homogenous density of the same background gas.

In an ongoing research project, we want to use a supersonic beam as an ultracold gas target in order to generate short rare gas ion pulses via femtosecond photoionization. For that application, we make use of the two unique characteristics of a supersonic expansion, namely the low perpendicular temperature and the high number density of the particles. The key factors limiting the achievable ion pulse duration are i) the randomly distributed thermal starting velocities of the generated photoions and ii) their repulsion in a propagating ion pulse due to Coulomb interaction. In order to reach an ion pulse duration of the order of one picosecond at ion energies in the keV range, the first limitation requires a temperature reduction of the gas target down to the $10^{-3}$ K regime. Temperatures of that order can be reached in a supersonic beam, if the generated photoions are extracted perpendicular to the beam propagation. The trick is that $T_\perp$ can be effectively controlled by a geometrical cooling effect, where the perpendicular component of the particle velocity is restricted by collimating the beam at large distance $x$ from its virtual source close the nozzle. Theoretical estimates reveal $T_\perp$ to scale inversely proportional with $x^2$. The drawback, however, is that the beam density also rapidly falls with increasing distance, and it is therefore critical to determine the number density of gas particles at a given distance $x$ and temperature $T_\perp$. Simulations of the resulting ion pulses reveal that the density of neutral gas atoms must reside in the range of about $10^{11}\,\text{cm}^{-3}$ in order to produce bunches consisting of 1-10 ions, which can then be accelerated and transported over distances of several millimeters without substantial space charge broadening.

As a prerequisite to these planned experiments, it is crucial to obtain knowledge of the beam properties, in particular of the nature (gas atoms vs. clusters), temperature and number density of the particles in the beam. We therefore characterize the beam in a twofold way. First, we retrieve information about the number density by ionizing the neutral gas particles using a VUV excimer laser and detecting the resulting ions using a linear time-of-flight (TOF) mass spectrometer. By comparison of the ions extracted from the supersonic beam to those generated in a homogenous background gas of the same species, we are able to calibrate the absolute number density of gas particles entrained in the beam. Second, we gather information about the beam density profile by scanning a skimmer in a plane perpendicular to the beam axis and measure the transmitted neutral particles via the signal of an ion gauge located behind the skimmer. Combining the results, we will compare them to theoretical predictions using the sudden freeze model (SFM) describing a supersonic expansion.

## 2. Experimental

The experimental setup consists of four differentially pumped vacuum chambers, separated by 2 or 3 skimmers respectively of varying diameter. The first chamber contains a pulsed piezoelectric gas valve (The Amsterdam Piezovalve [16]) equipped with a conical nozzle of $d = 150$ μm diameter and 40° opening angle, providing Argon gas pulses with a nominal pulse length of 5-200 μs and a repetition rate of up to 5 kHz at stagnation pressures in the range of 5 to 15 bar. The nozzle is mounted on a tiltable xy manipulator for beam alignment, and the gas load entering the expansion chamber is pumped by a turbomolecular pump with a nominal pumping speed of 1000 l/s for argon. A first skimmer of 1.5 mm aperture diameter is located at a distance of 125 mm from the nozzle, separating the expansion chamber from an intermediate chamber differentially pumped with a 300 l/s turbomolecular pump. A second skimmer of 1 mm aperture diameter located 225 mm downstream from the nozzle introduces the beam into a third transfer chamber pumped with a 300 l/s turbomolecular pump. All three chambers are manufactured according to ultrahigh vacuum standards and feature a base pressure < 10$^{-9}$ mbar with the supersonic beam switched off. Upon beam operation at a stagnation pressure of 5 bar, the pressure in the three chambers rises as a function of the (set) valve opening time as shown in Figure 1.

It is seen that the valve apparently starts to open at a set pulse width of 20 μs, followed by a linear increase up to about 100 μs. In this range, we therefore assume the actual pulse width $t_p$ to scale with the set valve opening time minus 20 μs. From the measured pressures, it is in principle possible to estimate the gas load delivered into different parts of the vacuum system. For the expansion chamber, the effective pumping speed is geometrically reduced to about 500 l/s for argon. From the pressure measurement at $t_p = 40\,\mu\text{s}$, this results in an average gas flow of about 1.8×10$^{16}$ argon atoms per pulse. Assuming a peaking factor of about 10 for the conical nozzle employed here (see below), this translates into 1.0×10$^{13}$ and 1.1×10$^{12}$ atoms going through the first and second skimmer, respectively.

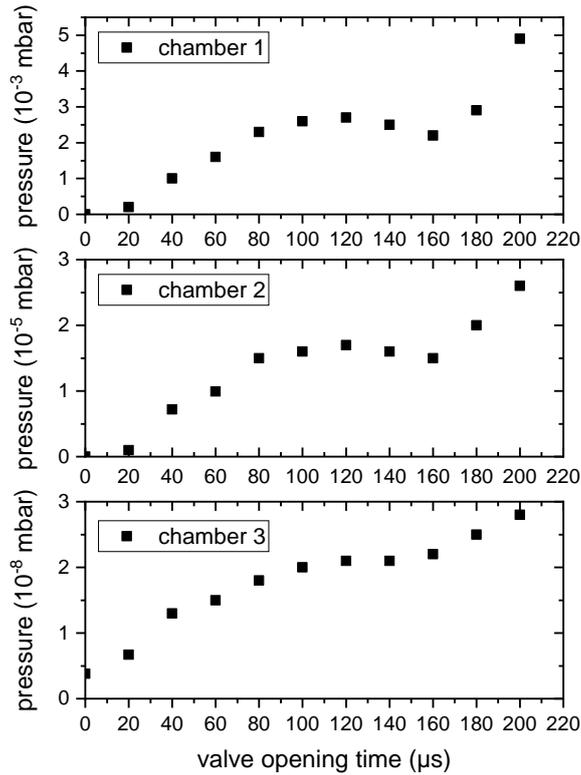

*Figure 1: Pressure in the three differential pumping stages vs. set valve opening time per pulse. The data were measured with argon at a backing pressure of 5 bar and a pulse repetition rate of 1 kHz.*

Analyzing the pressure measurement in chamber 3, however, we find a maximum possible beam-induced gas load of about $5.4 \times 10^{10}$ atoms entering this chamber, indicating that the combined transmission of the first and second skimmers must be less than 5%. A more detailed discussion of this observation will be given below. In any case, one can see that the three-stage differential pumping scheme employed here is suitable to ensure ultrahigh vacuum conditions during beam operation already in the third stage, a feature which is important in the context of the planned application of the generated supersonic beam.

The differential pumping system is followed by a fourth "analysis" chamber pumped by a 80 l/s turbomolecular pump, which contains a linear time-of-flight (TOF) spectrometer set up in Wiley-McLaren configuration [17]. The neutral gas particles are ionized using a pulsed laser beam directed perpendicular to the supersonic beam propagation as described below, and the resulting ions are extracted along the direction perpendicular to both the laser and particle beams and registered with a double microchannel plate (MCP) detector in chevron configuration. The TOF system allows a mass resolved detection of ionized gas particles and serves as a tool to investigate the density, composition and the velocity profile of the neutral beam. In normal operation, the analysis chamber is separated from the second differential pumping stage by a third skimmer of 1 mm aperture diameter, which is mounted on an xy manipulator and can be translated in the plane perpendicular to the beam propagation axis. In some experiments, however, this skimmer was dismounted. All skimmers feature sharp edges of 10 µm wall thickness.

A schematic of the experimental setup is depicted in Figure 2.

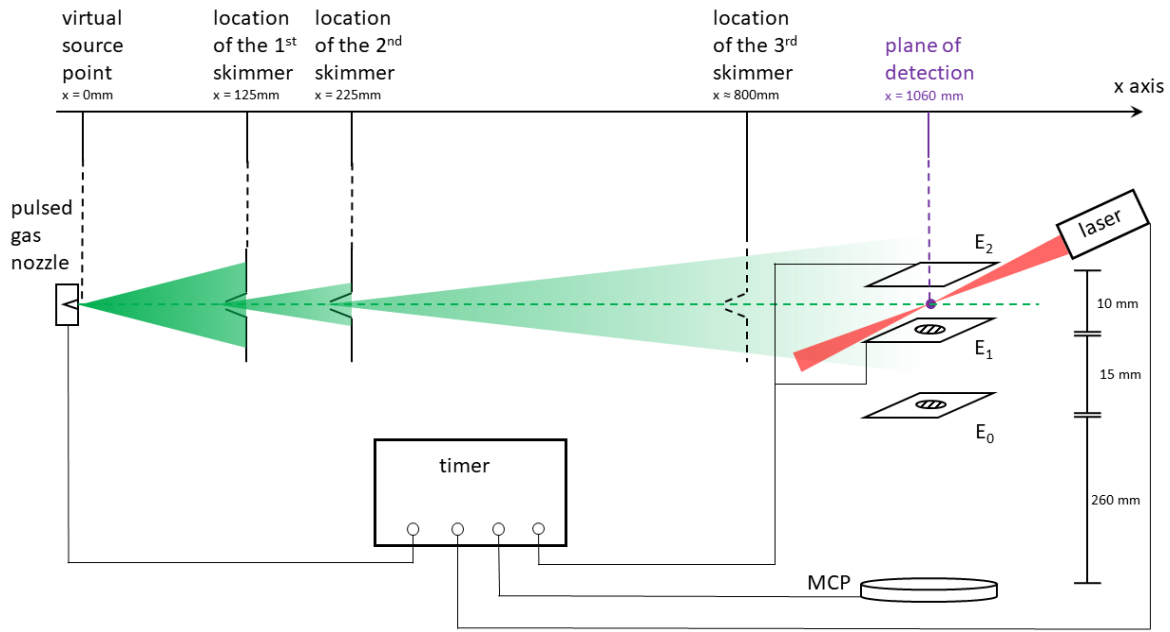

*Figure 2: Schematic of the experimental setup. A pulsed gas beam is skimmed down in diameter along the beam axis (x axis). At the plane of detection, a laser perpendicular to the beam axis generates electrons in the TOF volume, which are rapidly drawn onto the upper electrode $E_2$ and ionize the neutral gas atoms along their way. Those ions are then accelerated by the same electric field between the electrodes $E_1$ and $E_2$ perpendicular to the gas beam and the laser beam and focused onto a MCP detector via Wiley-McLaren configuration and can therefore be detected. For the timing of all experimental components a timer synchronizes the nozzle with the laser and the potentials of the TOF electrodes as well as the MCP.*

To adjust the beam axis up to the analysis chamber, the optical axis is coarsely aligned at first via a laser diode in place of the nozzle at atmosphere and then fine-tuned under vacuum conditions by maximizing the pressure in the individual chambers successively when the beam is running. Without the third skimmer in place, the second skimmer defines the beam shape at the plane of detection. A comparison of the opening diameter of the second skimmer (1 mm) with its distance from the nozzle (225 mm) yields a maximum tilt angle of 0.25° for the gas beam to pass through the skimmer, which is difficult to cover with the given manipulator. Thus, the beam alignment might not be optimal but sufficient to measure the beam properties at the plane of detection, as can be seen in Section 3.D.

Ionization of neutral gas particles is performed using an excimer laser (Coherent ExciStar XS 200) operated with an $F_2$/He gas mixture, thus providing pulsed VUV radiation with a wavelength of 157 nm (7.9 eV photon energy) at a maximum pulse energy of 3.0 mJ, a pulse length of 5-8 ns and a maximum repetition rate of 500 Hz. In principle, the laser radiation could be directly used to photoionize the neutral gas particles via non-resonant multiphoton absorption. For the case of argon as investigated here, simultaneous absorption of at least two photons is needed, which requires rather high photon flux densities in excess of $10^8$ W/cm² in order to be efficient. First attempts to measure the beam profile using that strategy, which were performed by scanning the tightly focused laser beam across the supersonic beam and monitoring the signal of photoionized Ar atoms as a function of laser focus position, proved unsuccessful. As will be described in detail below, it was found that the measured ion signal does not originate from direct photoionization of gas particles but is instead generated via *electron impact ionization* utilizing photoelectrons created by stray light from the pulsed laser radiation. These electrons are accelerated towards the two electrodes $E_1$ and $E_2$ of the TOF setup, which are set to positive potentials $\phi_1 = +1740\,\text{V}$ and $\phi_2 = +2040\,\text{V}$ in order to extract the ions onto the detector. In particular, electrons created at the surface of electrode $E_1$ are drawn to the upper electrode $E_2$, thereby gaining an energy of 300 eV which is suitable for efficient electron impact ionization. The argon ions as well as ionized residual gas atoms are then extracted along the direction perpendicular to both the supersonic and laser beam axes and detected by the MCP. A second electric field between the electrodes $E_1$ and $E_0$ (with $\phi_0 = 0\,\mathrm{V}$) allows to satisfy first order flight time focus conditions for the detected ions in order to increase the mass resolution. The synchronization between nozzle, laser and voltages is ensured by a Digital Delay Generator (Stanford Research Systems DG535) with two timing parameters: the laser delay and the nozzle delay. The former is the delay between the firing of the laser and the switching of the HV and is set to 60 ns, whereas the latter is the delay between firing the nozzle and laser pulses and lies in the order of hundreds of µs.

The signal measured by the MCP detector is processed by an electric circuit sketched in Figure 3.

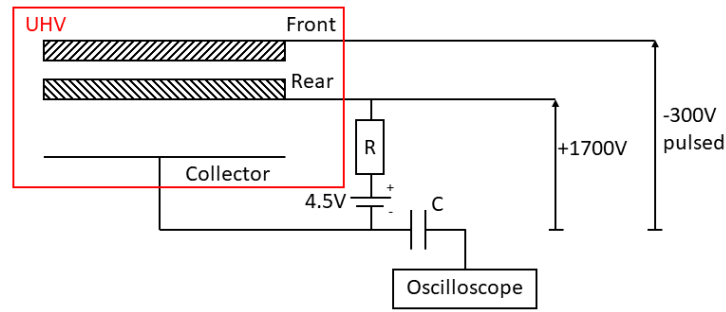

*Figure 3: Schematic of the electric circuit used to measure the signal. The rear of the MCP is floating on a voltage of +1700V while the front is gated with -300V. A resistor with R=1 kΩ and a battery pack providing 4.5V stabilize the output signal, while a capacitor with C=100 nF allows to extract the signal against the high voltage. It can then be displayed by an oscilloscope.*

Notably the front of the detector is also pulsed from ground potential to -300 V using a fast push-pull switch (Behlke HTS 31-GSM). Since the MCP gain varies by about three orders of magnitude between both states, the detector sensitivity is greatly reduced except when pulsed to -300 V. The detector pulse is synchronized with the TOF experiment in such a way that the gain is reduced during the firing of the laser pulse. This is necessary to ensure that the signal originating from scattered laser photons does not saturate the detector. During a time window of 5 µs around the expected ion arrival time, the maximum voltage of 2 kV is applied to the MCP, providing a sufficiently high gain for single ion detection using a transient digitizer with 1 ns time bin resolution. The spectra are averaged over 256 sweeps to minimize statistical errors as well as errors caused by electronic noise. Spectra of the residual gas were also measured, averaged over 256 sweeps and subtracted from the argon measurements. Finally, the resulting argon ion signal was smoothed with a linear moving average of 81 ns width. Such an exemplary spectrum is shown in Figure 4. Here, the largest peak at a flight time of 3.5 µs can be identified as $Ar^+$ ions, while the smaller peak at 2.55 µs can be identified as $Ar^{2+}$ ions. This presence of doubly charged Argon ions is another clear indicator for ion generation via electron impact ionization. A third peak at 4.9 µs corresponds to singly charged Argon dimer ions. Furthermore, the switching noise of the fast push-pull switch as well as some noise between the $Ar^{2+}$ and $Ar^+$ peaks are noticeable, but they do not interfere with the evaluation of the respective peaks, since only the time-integrated absolute areas of the measured flight time peaks are proportional to the number of ions created per pulse.

To calibrate the measured ion signal in terms of absolute number densities of the corresponding neutral particles, a gas inlet is mounted at the analysis chamber which allows to insert argon gas at a specific background pressure monitored by an ion gauge (Leybold IoniVac ITR 90) using a controlled leak valve. The comparison between the signals measured from the supersonic gas beam and the homogenous background gas then facilitates a quantitative determination of the number density in the beam, as described further in Section 3.D.

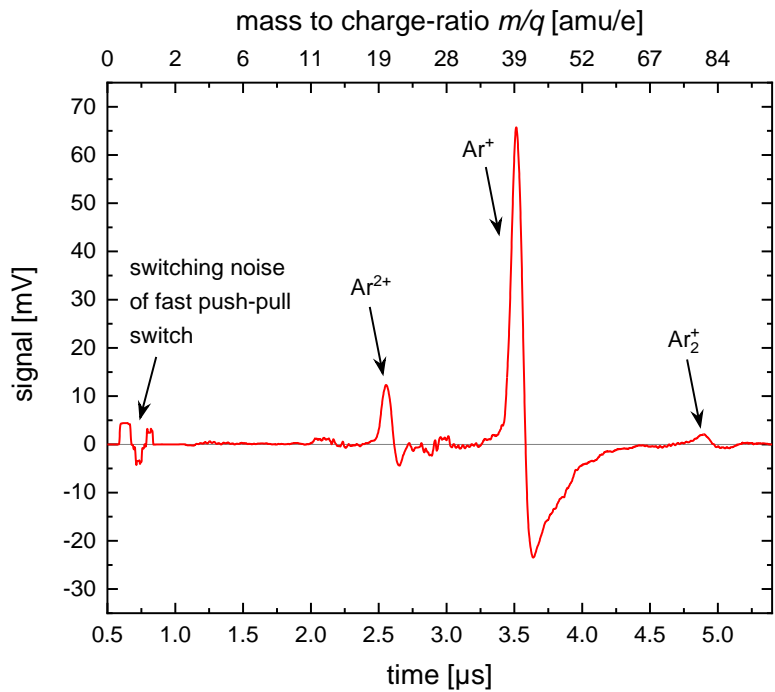

Figure 4: Typical time-of-flight spectrum averaged over 256 sweeps and smoothed with a 81 ns moving average. The time integration of a pulse yields a signal proportional to the number density in either the gas beam or the background gas.

## 3. Results and discussion
### A. Beam composition

Inspecting the time-of-flight mass spectrum shown in Figure 4, it is seen that the beam-induced Argon ion signal can be extracted with the procedure depicted above and that the singly charged ions at mass 40 can clearly be distinguished from the doubly charged ions. Moreover, a non-vanishing peak at mass 80 is assigned to the Argon dimer, indicating that the fraction of Argon clusters in the beam is small. According to Hagena [18], a good scaling parameter for cluster formation is given by $c = p_0 R_n^q T_0^{-r}$, $p_0$ being the stagnation pressure, $R_n$ the nozzle radius, $T_0$ the nozzle temperature and $q$ and $r$ two parameters. For Argon, cluster formation should be negligible if $c$ is smaller than $c_{max,Ar} = 2.6 \times 10^{-7}$ mbar m$^{0.88}$ K$^{-2.3}$ [9]. Inserting $p_0 = 5$ bar, $R_n = 75$ µm and $T_0 = 300$ K, we find $c = 8.8 \times c_{max,Ar}$, so that cluster formation should in principle be possible. Comparing the integrated areas of each peak, we find that the ratio of the Argon dimer makes up only 1 % of the beam. However, we presume that driving the nozzle with a higher stagnation pressure up to 15 bar should introduce a higher probability for the formation of Argon dimers and larger clusters, which could also be used to generate bunches of multiatomic ions using the supersonic beam.

### B. Parallel velocity and speed ratio

The spread in parallel velocity of the particles entrained in the beam corresponds directly to its parallel temperature and is therefore a crucial characteristic of the beam itself. It can be described by the speed ratio *S*, a dimensionless number which denotes the ratio between the terminal parallel particle velocity $v_{\|,\infty}$ and the full width half maximum (FWHM$_v$) of its distribution, multiplied by a factor $2\sqrt{\ln(2)}$. In order to gain information on both quantities, the ionization laser was placed in the center of the TOF extraction region and the ion signal was recorded as a function of the delay between the firing of the nozzle valve and the laser pulse. Figure 5 shows the resulting variation of the ion signal. The data were taken at a backing pressure of 5 bar argon gas at room temperature with a nominal nozzle opening time of 50 μs and a pulse repetition rate of 500 Hz. Due to the long distance of 1060 mm between the nozzle and the ionization region, analysis of these data reveals information about the average flow velocity as well as its thermal distribution. In that context, it is of note that the measured delay time distribution represents a convolution between the spread of the parallel particle velocity and the temporal width of the beam pulse generated by the nozzle valve. In cases where the beam is intersected at relatively small distance from the nozzle, the delay measurement therefore primarily delivers information about the temporal pulse shape. Using a similar piezo valve as employed here, Irimia et al. [19] have measured the pulse shape generated for different seeded beams (2 % NO$_2$ and 0.5% NO) in Helium at short distance (~ 100 mm), where the pulse broadening due to the beam velocity spread is almost negligible. For a nominal pulse width (i.e., the width of the electric pulse driving the piezo valve) of 23 μs duration, they found actual gas pulse widths of 12 and 24 μs (FWHM), respectively. Based on these data, along with our pressure measurement described above, we assume the actual temporal pulse duration to be shorter than the nominal pulse width. For the setting of 50 μs applied here, the pulse width should therefore be significantly smaller than the width of the measured delay curve displayed in Figure 5, so that – due to the long distance from the nozzle used in our experiments – the delay time distribution measured here reveals mixed information regarding the temporal beam profile and the particle velocity spread.

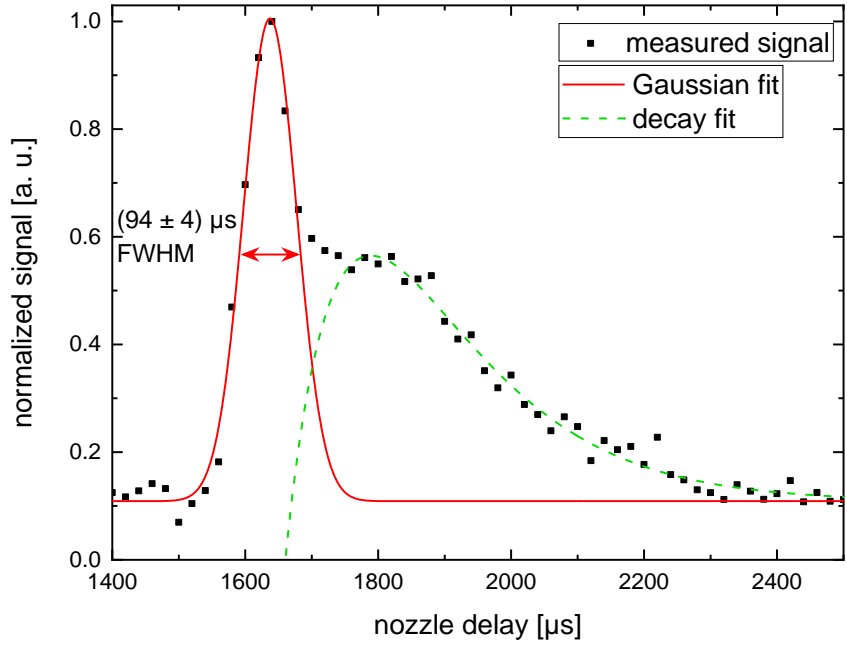

*Figure 5: Measured signal as a function of the delay between the firing of the nozzle valve pulse and the ionization laser pulse. The nozzle valve was operated with a nominal pulse width of 50 µs, a backing pressure of 5 bar and a repetition rate of 500 Hz.*

A first inspection of the data reveals an asymmetric pulse shape with a width of about 145 µs (FWHM). In principle, the asymmetric shape could indicate either a non-thermal velocity distribution or an asymmetric temporal pulse shape. In view of the fact that symmetric velocity distributions are both theoretically expected [5] and have been measured before [8, 19], we believe that the asymmetric shape is caused by a tail in the temporal pulse shape. In fact, asymmetric pulses featuring either tails, a shoulder or even a second maximum at longer times have been observed with a piezoelectric nozzle valve of the same kind (and manufacturer) as applied here [19]. We therefore interpret the data presented in Figure 5 as follows. In order to evaluate the thermal velocity spread, we fit a Gaussian to the data points up to delay times of about 1670 µs as indicated by the solid line in Figure 5. From the maximum of the resulting fitting curve, we deduce a mean flight time of 1640 µs corresponding to a mean flow velocity of $v_{\|,\infty} = 645\,\text{m/s}$. Furthermore, to correct the FWHM$_t$ of (94 ± 4) µs of the fitting curve, we crudely approximate the gas pulse to be rectangularly shaped with the set nominal pulse width of 50 µs minus 20 µs. Then, a simple calculation yields a deconvoluted thermal spread of FWHM$_{t,dec}$ = (92 ± 4) µs, from which the speed ratio can be determined as

$$S = \frac{2\sqrt{\ln(2)}\,v}{\text{FWHM}_v} = \frac{2\sqrt{\ln(2)}\,t}{\text{FWHM}_{t,dec}} = 2\sqrt{\ln(2)}\,\frac{1640\,\mu\text{s}}{(92\pm 4)\,\mu\text{s}} \approx 30 \pm 1.3. \qquad (3.1)$$

This value is significantly smaller than the theoretical value obtained, for instance, from the so-called "sudden freeze model" (SFM, see below) [20]

$$S_\infty = A\left[\sqrt{2}n_0 d\left(\frac{53C_6}{k_B T_0}\right)^{1/3}\right]^B \tag{3.2}$$

with constants $A = 0.527$, $B = 0.545$ and $C_6/k_B = 41.2\times 10^{-55}\,\text{Km}^6$ as reported by Schofield et al. [14]. The nozzle density $n_0 = 1.2\times 10^{20}\,\text{cm}^{-3}$ (determined by the stagnation pressure $p_0 = 5\,\text{bar}$ at temperature $T_0 = 300\,\text{K}$) along with the nozzle diameter $d = 150\,\mu\text{m}$ yields $S_\infty \simeq 125$. The large difference between experimental value and theoretical estimate is not unexpected and presumably arises from the fact that temporary cluster formation during the jet expansion as well as influences of the skimmers limit the achievable speed ratio. In fact, it was shown by Hillenkamp et al. [8] that for a conical nozzle (with a smaller opening angle and nozzle diameter than used here) the speed ratio for Argon plateaus at around $S \approx 30$, rather independent of the stagnation pressure. Based on these data, , a speed ratio of $S \approx 30$ for our experimental setup seems plausible. This converts to a parallel temperature

$$T_{\|,\infty} = \frac{m}{2k_B}\left(\frac{v_{\|,\infty}}{S}\right)^2 = (1.1 \pm 0.1)\,\text{K} \tag{3.3}$$

which seems plausible as well in comparison to work from Even [5] and coworkers [8].

The question remains what causes the rather long tail in the signal for increasing nozzle delays, which is fitted in Figure 5 (green dashed line) with a somewhat linear rise and a subsequent exponential decay. In principle, such a behavior indicates that the piezo valve does not close properly inside the nozzle, maybe due to the relatively high repetition rate used in our experiment. In that case, the volume in the analysis chamber would be steadily filled with gas atoms for some time after the pulse (leading to the linear signal increase), which then get pumped out (leading to the exponential signal decay). It should be noted that similar tails have been observed as well in ref. [19], where it was found that these effects become more pronounced at high repetition rates.

### C. Background gas measurement

In order to obtain an absolute calibration of the beam density, we compare the ion signal determined for the supersonic beam with that measured for a homogenous atom density. For that purpose, we backfill the analysis chamber with argon gas to a certain partial pressure. The laser focus is then positioned at the center of the TOF extraction volume and the argon ion signal resulting from the same photoelectron ionization process is measured as a function of the Argon partial pressure. As indicated by Meng et al. [15], processes like ion scattering and charge exchanges [21] can contribute to the signal in a nonlinear manner, so that the linearity of the measured ion signal with the argon pressure in the chamber should be checked. Figure 6 shows the results, which indicate a clearly linear behavior in the regime between 1 to $12\times 10^{-7}\,\text{mbar}$, which is of interest here.

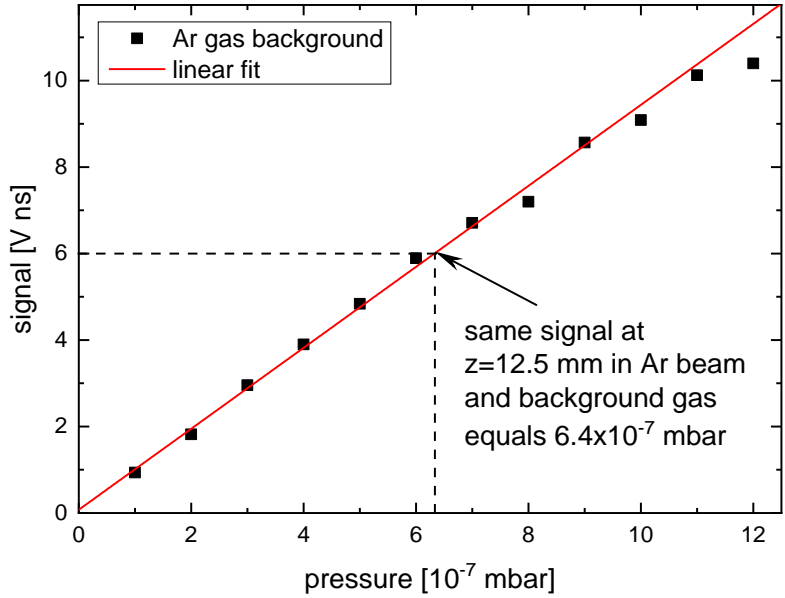

*Figure 6: Signal of argon ions measured with the analysis chamber homogenously backfilled with argon gas vs. partial argon pressure in the chamber. The ionization laser was operated at 30 Hz repetition rate and 3.0 mJ energy per pulse and positioned in the center of the interaction region.*

The linear fit yields the identical time-integrated signal for the beam and bulk for a bulk pressure of $6.4 \times 10^{-7}$ mbar, which converts to a number density of $n_{Bulk} = 1.24 \times 10^{10}$ cm$^{-3}$ via $n = f \cdot p / k_B T_0$, $k_B$ being the Boltzmann constant and $T_0 = 300$ K. The factor $f$ covers a correction factor for the pressure measurement which depends on the ion gauge and is specified by the manufacturer to be 0.8 for Argon. A calculation of the ionization volume then allows to determine the number of Argon atoms hitting the detector, which is done in the following.

The electrodes $E_1$ and $E_0$ each contain a grid (transmission: 0.9) with a diameter of 12.5 mm, limiting the ion extraction volume, i.e., the volume from which ions can be extracted and hit the detector, to a cylinder of the same diameter. Flight-time simulations performed with SIMION 8.0 show that this volume is further reduced to an effective diameter of $D_{eff}$ = 6 mm, which represents an average over all diameters along the extraction direction from which ions can hit the detector. Since the electric field between the electrodes breaks out for larger diameters, ions created outside of that cylinder are therefore accelerated onto the non-transmitting part of the electrodes. Considering the distance of 10 mm between the electrodes $E_2$ and $E_1$ as well, the ion extraction volume can be calculated as 280 mm³, resulting in a total number of detectable particles of $N_{Bulk} = 3.5 \times 10^9$ per laser shot on average.

The laser position was then varied along the $z$-axis and the signal was measured at each laser position for the beam and for the bulk, respectively. Figure 7 shows the result, which reveals no significant difference between the supersonic beam and the bulk background gas signal for any given laser position. At first sight, this result appears surprising. For the bulk background gas, the number density of Argon atoms is constant and the signal variation along the $z$-axis should therefore mimic the imaging characteristics of the TOF spectrometer. The identical curve measured for the supersonic beam, however, renders this interpretation questionable. It would only hold if the beam diameter is broad enough to ensure a constant atom density across the entire ionization volume of 10 mm height. As shown in the following section, this is not the case. We therefore interpret the data shown in Figure 7 in a different way. In fact, the apparent insensitivity of the measured signal from the position of the laser beam provides another indication that the detected ions cannot be generated by direct photoionization. If the signal is produced by photoelectron ionization, on the other hand, the curves displayed in Figure 7 simply reflect the efficiency of photoelectron production as a function of the laser beam position. Since the resulting photoelectron ionization efficiency does not depend on the number density distribution of the neutral atoms, this would naturally explain the identical curves.

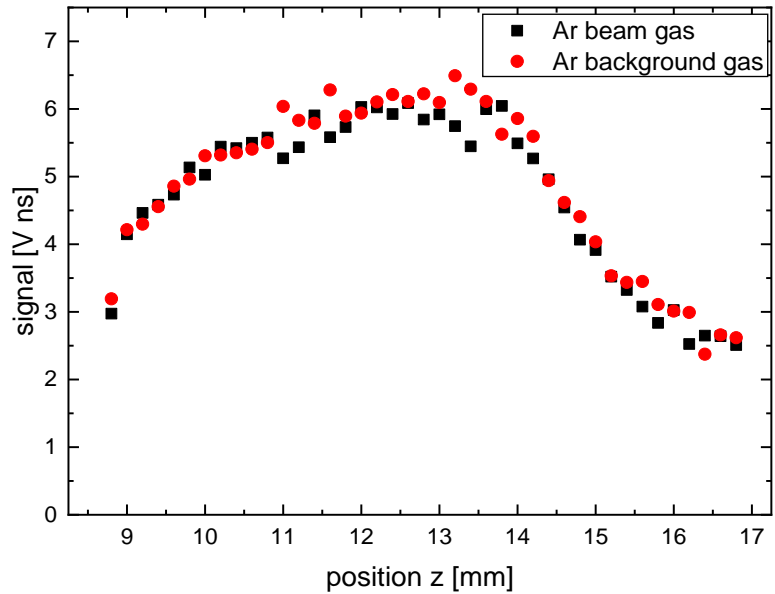

Figure 7: Comparison of the measured signals for the supersonic Argon gas beam and the bulk Argon background gas as a function of the ionization laser position along the ion extraction axis.

### D. Beam profile measurement and particle density in the beam

In order to gain information about the beam profile, a third skimmer with a diameter of 1 mm is introduced into the beam line and mounted onto an xy-manipulator at a distance of about 800 mm from the nozzle (see Figure 2). Furthermore, the whole TOF setup is removed and an ion gauge is placed at the end of the beam line. By scanning the skimmer position perpendicular to the beam and monitoring the measured ion gauge signal we can therefore determine the cross sectional profile of the beam. Figure 8 displays a sketch of the experimental setup.

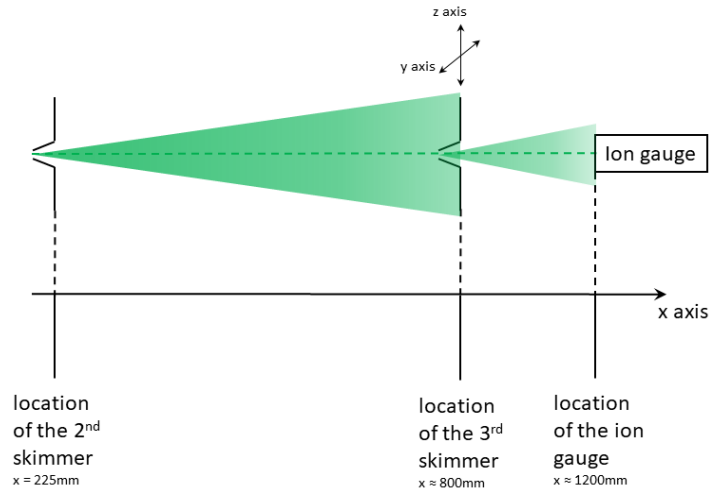

Figure 8: Experimental setup for the measurement of the beam profile. The TOF is removed and a third skimmer with a diameter of 1 mm is introduced in the beam line. Scanning this skimmer along the y and z axis allows to monitor the pressure measured by an ion gauge (Leybold IoniVac) further downstream of the beam.

The nozzle was run at 500 Hz with an opening time of 100 µs and a backing pressure of 10 bar against a background pressure of $4.8 \times 10^{-8}$ mbar, which was subtracted from the measured pressures when the beam was running. The center beam pressure was measured to be $2.83 \times 10^{-7}$ mbar, which is significantly higher than the background pressure and therefore allows to measure the beam profile this way. The normalized ion gauge signal as a function of the skimmer position along the z axis is shown in Figure 9 and yields a peak profile with falling edges from the center with a total $FWHM_z$ of 4.75 mm.

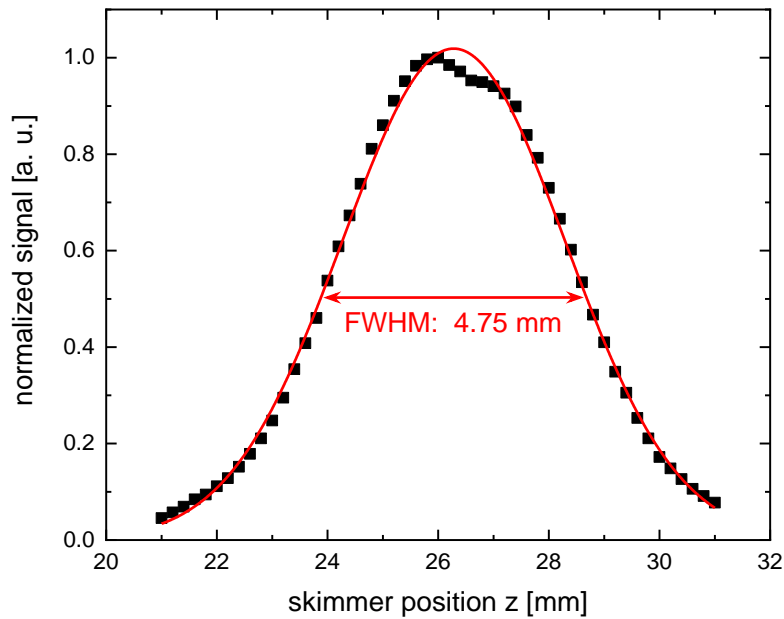

Figure 9: Normalized pressure signal of the beam as a function of the 3rd skimmer position along the z axis. The Gaussian fit curve describes the measured data points rather well, save for a slight deviation at the peak top. The fit of the peak yields a FWHM of 4.75 mm.

The general shape of the beam consists of two coaxial Gaussians of the form $f_z(z) = \sum_{i=1,2} c_{z,i} \exp(-z^2/2\sigma_{z,i}^2)$ with the boundary condition $c_{z,1} + c_{z,2} = 1$, as often used in the literature [7, 22, 23] to describe two contributions of a virtual source. However, since the profile is measured far away from the nozzle, we can safely neglect the so called *warmer* source, implying that only the narrow, *colder* source can contribute to the signal in the center of the beam:

$$f_z(z) = \exp\left(-\frac{z^2}{2\sigma_z^2}\right) \quad (3.4)$$

Projecting the opening diameter of the second skimmer to the position of the probing third skimmer geometrically, we estimate a width of about 4.18 mm, which is in fair agreement with the measured FWHM if we take the finite diameter of the third skimmer into account as well. A small deviation from the Gaussian fit at the peak top is likely caused by not perfectly aligned skimmers in the beam line, as described above in the experimental section. This effect is even more noticeable in the measurement along the y axis, resulting in a slightly smaller FWHM$_y$ of 4.6 mm, though with a more asymmetric peak form. Nevertheless, we take both measured standard deviations $\sigma_{y,z}$ for further discussion, since they allow us to perform further calculations analytically.

We can then give a simple estimation of the perpendicular temperature by associating the standard deviation of the profile function with the appropriate Boltzmann distribution for $T_\perp$ as

$$\exp\left(-\frac{mv_\perp^2}{2k_B T_\perp}\right) = \exp\left(-\frac{z^2}{2\sigma_z^2}\right). \quad (3.5)$$

Considering the geometric relation $v_\perp/v_{\parallel,\infty} = z/\Delta x$, $\Delta x$ being the flight distance for the Argon atoms from the nozzle to the ion gauge, we can estimate the perpendicular temperature as

$$T_\perp = \frac{mv_{\parallel,\infty}^2}{k_B} \cdot \left(\frac{\sigma_z}{\Delta x}\right)^2 \quad (3.6)$$

Inserting the measured values for $v_{\parallel,\infty}$, $\sigma_z$ and $\Delta x \approx 1200\,\text{mm}$, this yields a perpendicular temperature $T_\perp$ of 5 mK.

We now evaluate the center line density $\rho_0$ from the measurements depicted in Figure 5 and Figure 9. The total number of atoms in the beam is given by the requirement

$$N_{Bulk} = N_{Beam} = \iiint \rho(x,y,z)\,dV, \quad (3.7)$$

with the density function of a gas pulse in Cartesian coordinates

$$\rho(x,y,z) = \rho_0 f_x(x) f_y(y) f_z(z), \quad (3.8)$$

$f_{x,y,z}$ being the profile function in the appropriate directions. For the perpendicular directions $y$ and $z$, the single Gaussian distribution functions are used, according to Equation (3.4). The distribution $f_x(x)$ along the $x$-axis parallel to the beam is also Gaussian-shaped with a standard deviation of $\sigma_x = v_{\parallel,\infty} \sigma_t = 26\,\text{mm}$, as could be extracted from Figure 5. For the present estimate, however, we assume the profile to remain constant along the length of the diameter $D$ of the cylindrical ionization volume, which is justified by the condition $\sigma_x \gg D$. Here, the diameter $D$ is averaged within the measured FWHM$_z$ of 4.75 mm around the center of the extraction axis instead of the whole extraction volume, since the beam is more constrained to the center of the TOF volume in contrast to the bulk gas, which fills up the whole ionization volume homogeneously. Therefore, an average of $D = 2r \approx 5$ mm for the center of the ionization volume is justified by the same simulation performed with SIMION 8.0 in Section 3.C. Furthermore, since the diameter and the standard deviation $\sigma_y$ lie in comparable dimensions, we have to constrain the integration limit along one axis ($x$ or $y$) via $x^2 + y^2 = r^2$. We choose the constrain along the x axis for the simplification of the integration along the $x$-axis. Lastly, we raise the integration limits for the $z$-direction to $\pm\infty$ without significantly changing the result, since we can safely consider the standard deviation $\sigma_z$ to be much smaller than the distance between the electrodes limiting the TOF volume along the $z$-direction. Therefore, we can make use of the standard integral

$$\int_{-\infty}^{+\infty} \exp\left(-\frac{\xi^2}{2a^2}\right) d\xi = \sqrt{2\pi} \cdot a \tag{3.9}$$

to solve the integral in Equation (3.7) along the $z$-direction, which has now the form

$$N_{Beam} = \sqrt{2\pi} \cdot \sigma_z \rho_0 \int_{-r}^{+r} \int_{-\sqrt{r^2-y^2}}^{+\sqrt{r^2-y^2}} dx\, f_y(y)\, dy. \tag{3.10}$$

Considering $N_{Bulk} = N_{Beam}$ as well as the pulse shape $f_y$ from Equation (3.4) and subsequently inserting the measured values for $\sigma_{y,z}$ respectively, a quick calculation with *Wolfram Mathematica* yields the resulting center line density to $\rho_0 = (4.5 \pm 0.12) \times 10^{10}\,\text{cm}^{-3}$.

We can compare this value with an estimation of the average gas load per pulse, as already proposed by Meng et al. [15]. Since we know the central Argon partial pressure to be $1.88 \times 10^{-7}$ mbar at 500 Hz repetition rate (considering the correction for the residual gas pressure and the factor *f*) and the pumping speed of the turbomolecular pump in the analysis chamber (80 l s$^{-1}$), we can calculate the average gas load as $1.5 \times 10^{-8}$ bar l s$^{-1}$. This converts to $3.0 \times 10^{-11}$ bar l per pulse, which conforms to $N_0 = 7.2 \times 10^{11}$ particles per pulse. We can now estimate the center line density via

$$\rho = \frac{1}{\left(\sqrt{2\pi}\right)^3} \cdot \frac{N_0}{\sigma_x \sigma_y \sigma_z}, \tag{3.11}$$

$\sigma_{y,z}$ being the measured standard deviations from the profile measurements along the $y$- and $z$-axes and $\sigma_x$ the standard deviation along the beam axis. This results in an ideal center line density of $\rho = 4.5 \times 10^{11} \, \text{cm}^{-3}$, however, since we need to cover the long tail shown in Figure 5, we assume the temporal width $\sigma_t$ here to be around 5 times larger than the Gaussian fit. This factor is owed to the comparison of the total area under the data points with the area under the curve of the Gaussian fit alone, reducing the density to rather $\rho \approx 1.0 \times 10^{11} \, \text{cm}^{-3}$. Moreover, the average gas load calculation greatly depends on the pumping speed of the turbomolecular pump, which for Argon most likely differs from the nominal pumping speed of 80 l s⁻¹, reducing the center line density even further. Therefore, it is quite remarkable to have the center line density measured via photoelectron ionization to agree within a factor 2 with the gas load estimation.

We can further compare the measured center line density with the theoretical value predicted by the sudden freeze model (SFM) developed by Beijerinck and Verster [9]. For that purpose, we briefly review the SFM itself and derive some values from it for our source, allowing to compare with similar jet expansion measurements. Within the SFM, the supersonic gas expansion is separated into two regimes. The continuum flow regime covers the adiabatic expansion close to the nozzle and can be described as a gas with colliding molecules in full thermal equilibrium. Up to the so-called freezing distance, the collision frequency decreases rapidly and the trajectories can be considered as straight lines afterwards with no collisions at all. In this molecular flow regime, standard definitions of temperature fail, leading to a decomposition into a parallel temperature $T_\parallel$ and a perpendicular temperature $T_\perp$. While $T_\parallel$ remains constant along the expansion axis (the $x$-axis in our experiment), $T_\perp$ diminishes according to

$$T_\perp(x) = T_\perp(x_F) \left(\frac{x}{x_F}\right)^{-2} \tag{3.12}$$

for $x > x_F$ ($x_F$ being the freezing distance). This decrease is a purely geometric effect, where no heat transfer occurs, since the straight trajectories can be traced back independently to a so-called virtual source point with a constant virtual source radius

$$R = \frac{\alpha_\perp(x)}{v_{\parallel,\infty}} x = \sqrt{\frac{2 k_B T_\perp(x)}{m}} \cdot \frac{x}{v_{\parallel,\infty}} \tag{3.13}$$

implying the relation $T_\perp(x) \sim x^{-2}$ already mentioned above. At the freezing distance $x = x_F$ the parallel temperature equals the perpendicular temperature, which allows to calculate the virtual source radius. The freezing distance can be approximated according to

$$x_F \approx d \left(\frac{S_\parallel}{C_1} \sqrt{\frac{2}{\gamma}}\right)^{\frac{1}{\gamma-1}}, \tag{3.14}$$

With $C_1 = 3.232$ and $\gamma = 5/3$ for a monoatomic gas [14], which results in a freezing distance of $x_F \approx 5\,\text{mm}$. Therefore, the virtual source radius can be calculated with Equation (3.13) to $R = 165\,\mu\text{m}$ and the perpendicular temperature at the position of the ion gauge can be determined via Equation (3.12) to $T_\perp(x = 1200\,\text{mm}) = 0.02\,\text{mK}$. This is clearly lower than the temperature estimated from the beam profile measurement by more than two orders of magnitude, however the theoretical temperature is calculated for an unskimmed expansion, neglecting the heating effect of the skimmers in the beam. Simulations show an increase in density shortly after each skimmer entrance, resulting in colliding gas atoms and thus heating of the beam perpendicular to its propagation axis by an order of magnitude at least [5]. Therefore, it is quite acceptable for the parallel temperature in the beam to be that much higher.

The sudden freeze model further predicts the center line density of the beam to decrease according to

$$n(x) = n_0 \left(\frac{x}{x_{ref}}\right)^{-2}, \qquad (3.15)$$

with $x_{ref} = R_n / a(\gamma) = 60\,\mu\text{m}$ being a reference distance given by the nozzle radius $R_n$ and a dimensionless constant $a(\gamma) = 0.808$ for a monoatomic gas. For the flight distance of 1060 mm between nozzle exit and the plane of detection, Equation (3.15) yields a number density of $n(x = 1060\,\text{mm}) = 3.9 \times 10^{11}\,\text{cm}^{-3}$ for a backing pressure of 5 bar, which lies one order of magnitude above the measured number density. However, one has to take into account that the calculations within the SFM are performed with a sonic nozzle, while the nozzle we use is conical with a full opening angle of 40°. As has been shown by Even [5], such a difference in nozzle shape can increase the center line density by more than one order of magnitude. Therefore, the difference between theoretical and experimental center line density is likely more than one order of magnitude, a finding which has been reported before [14, 15]. This huge discrepancy is presumably due to the presence of the skimmers, especially the second one which defines mostly the beam shape and number density. While the center line density is calculated for an unskimmed expansion, clogging of the skimmers can drastically reduce their transmission characteristics and therefore the density in the beam. We can compare the center line density at the location of the second skimmer entrance ($x \approx 200\,\text{mm}$) with simulations performed by Luria et al. [6], who report an expected number density of $2 \times 10^{22}\,\text{m}^{-3}$ for a nozzle diameter of 200 µm with a conical 40° opening angle and after a flight distance of 200 mm as well, though with a Helium beam at 30 bar backing pressure. Scaling the number density back to our measurements with 5 bar, we find an estimated number density of $3 \times 10^{21}\,\text{m}^{-3}$ at the skimmer entrance, and the simulations exhibit a transmission of less than 10% for a 1 mm diameter skimmer at these beam densities [5]. Even a decrease of the density by an order of magnitude raises the transmission only slightly, therefore it is safe to say that the second skimmer reduces the ideal center line density by more than an order of magnitude.

## 4. Conclusion

Using laser induced photoelectron ionization of neutral gas atoms, we have measured the particle density within a pulsed supersonic argon beam. Obtaining knowledge regarding the shape of a single gas pulse allowed to determine the peak center line density by comparison to a known density of a bulk gas, yielding a value of $\rho_0 = (4.5 \pm 0.12) \times 10^{10}\, cm^{-3}$. This density was shown to be in accordance with other direct measurements as well as theoretical predictions and computer simulations, attributing the discrepancy between theoretical considerations and experimental data to the well-known skimmer interference effects. Furthermore, we were able to measure the parallel temperature to $T_{\|,\infty} = (1.1 \pm 0.1)\,\text{K}$ and estimate the perpendicular temperature to be in the mK regime, which is a reasonable result compared to other measurements and theoretical predictions as well. With the opening time of the piezovalve and the backing pressure of the nozzle, we have two parameters which allow us to control the shape and density of the gas pulses even further. Therefore, we are quite encouraged to make use of the supersonic jet expansion described here with Argon gas to generate ultra-short ion pulses using the concept described in the introduction.

## Acknowledgements

The authors gratefully acknowledge financial support from the Deutsche Forschungsgemeinschaft (DFG, German Research Foundation) - Projektnummer 278162697 - SFB 1242.